\RequirePackage{fix-cm}
\documentclass[twocolumn]{svjour3}          
\smartqed  
\usepackage{graphicx,rotate}
\usepackage{epstopdf}
\usepackage{hyperref}
\usepackage{subfigure}
\usepackage[titletoc]{appendix}
\usepackage{textgreek}
\usepackage{amsmath,amssymb}

\begin{document}

\title{Micro-Macro Transition and Simplified Contact Models for Wet Granular Materials}


\author{Sudeshna Roy \and	 Abhinendra Singh \and Stefan Luding \and Thomas Weinhart
}


\institute{S. Roy \at
              Multi Scale Mechanics (MSM), Engineering Technology \\
              (CTW) and MESA+, University of Twente\\
              P.O.Box 217, 7500 AE Enschede, The Netherlands\\
              Tel.: +31-(0)53 489-3301\\
              \email{s.roy@utwente.nl}           
           \and
           A. Singh \at
              Multi Scale Mechanics (MSM), Engineering Technology \\
              (CTW) and MESA+, University of Twente\\
              P.O.Box 217, 7500 AE Enschede, The Netherlands\\
              Tel.: +31-(0)53 489-2694\\
              \email{a.singh@utwente.nl}           
           \and
           S. Luding \at
              Multi Scale Mechanics (MSM), Engineering Technology \\
              (CTW) and MESA+, University of Twente\\
              P.O.Box 217, 7500 AE Enschede, The Netherlands\\
              Tel.: +31-(0)53 489-4212\\
              \email{s.luding@utwente.nl}           
           \and
           T. Weinhart \at
              Multi Scale Mechanics (MSM), Engineering Technology \\
              (CTW) and MESA+, University of Twente\\
              P.O.Box 217, 7500 AE Enschede, The Netherlands\\
              Tel.: +31-(0)53 489-3301\\
              \email{t.weinhart@utwente.nl}           
}

\date{Received: date / Accepted: date}

\maketitle

\begin{abstract}
Wet granular materials in a quasi-static \\ steady state shear flow have been studied with discrete particle simulations. Macroscopic quantities, consistent with the conservation laws of continuum theory, are obtained by time averaging and spatial coarse-graining. Initial studies involve understanding the effect of liquid content and liquid properties like the surface tension on the macroscopic quantities. Two parameters of the liquid bridge contact model have been studied as the constitutive parameters that define the structure of this model (i) the rupture distance of the liquid bridge model, which is proportional to the liquid content, and (ii) the maximum adhesive force, as controlled by the surface tension of the liquid. Subsequently a correlation is developed between these micro parameters and the steady state cohesion in the limit of zero confining pressure. Furthermore, as second result, the macroscopic torque measured at the walls, which is an experimentally accessible parameter, is predicted from our simulation results as a dependence on the micro-parameters. Finally, the steady state cohesion of a realistic non-linear liquid bridge contact model scales well with the steady state cohesion for a simpler linearized irreversible contact model with the same maximum adhesive force and equal energy dissipated per contact.
\end{abstract}

\section{Introduction}
\label{intro}
Granular media are collections of microscopic grains having athermal interactions through dissipative, frictional or cohesive contact forces. External force leads to granular flow under the condition of applied shear stress exceeding the yield shear stress. After a finite shear strain, at constant rate, a steady state establishes with a typically lower shear stress, depending on both 
strain rate and pressure \cite{singh2014effect}. Most studies in granular physics focus on dry granular materials and their flow rheology. However, wet granular materials are ubiquitous in geology and many real world applications where interstitial liquid is present between the grains. Simplified models for capillary clusters \cite{mitarai2009simple,ulrich2009cooling} and wet granular gases \cite{strauch2012wet} were introduced before. The rheology of flow for dense suspension of non-Brownian particles have been studied in Ref. \cite{huang2005flow,huang2007viscosity,bonnoit2010mesoscopic}. We study the local rheology of weakly wetted granular materials in the quasistatic regime with the Discrete Element Method (DEM) using the open-source package MercuryDPM \cite{so89272,thornton2012modeling} in a shear cell set-up, where the relative motion is confined to particles in a narrow region away from the walls, called shear band \cite{schall2010shear,howell1999stress}. We study partially saturated systems in the pendular regime, with a very low level of water content, where the formation of liquid bridges between particle pairs leads to development of microscopic tensile forces. This tensile forces generated at particle level results in cohesion at macroscopic scale. Earlier studies have been done for liquid bridge in the pendular regime to understand the effect of liquid bridge volume and contact angle on different macroscopic quantities like the steady state cohesion, torque and shear band properties \cite{schwarze2013rheology,gladkyy2014comparison,herminghaus2005dynamics,richefeu2007shear,darabi2010modeling}. Other studies for unsaturated granular media observe fluid depletion in shear bands \cite{mani2012fluid,mani2013liquid}. However, there is no theoretical framework or concrete model available yet that defines the exact correlation between the micro parameters like the liquid bridge volume and the surface tension of the liquid with the steady state cohesion.
\par
In order to develop a micro-macro correlation for the liquid bridge contact model, we initially study the structure of the micro contact model. How is the structure of the liquid bridge contact model affected by the microscopic parameters? How does this influence the steady state cohesion? Here we study in detail on the effect of these parameters on the macro results. For example, the effect of maximum interaction distance, or the distance at which the liquid bridge between two interacting particles ruptures, is studied by varying the liquid content. On the other hand other parameters like surface tension of the liquid and contact angle affect the magnitude of force acting between the particles when in contact \cite{herminghaus2005dynamics,willett2000capillary}. Various surface tension of liquids give a large scale variation of the capillary force and this allows us to study the effect of maximum force on the macroscopic properties. Furthermore, in the consecutive analysis, we re-obtain the macro-rheology results in the shear band center from the torque, torque being an experimentally measurable quantity.
\par
The liquid bridge interactions between the particles are defined by the free-surface equilibrium shapes and stability of the bridge configuration between them \cite{alencar2006monte,lubarda2014stability,soulie2006influence}. Phenomenologically, even the simplified models of liquid bridges are quite complex in nature. In order to improve the computational efficiency for wet granular materials, we replace the non-linear interactions of liquid bridges with a simpler linear one. But in what way can a non-linear model like the liquid bridge contact model be replaced by a linear model? When can we say that the two different contact models are analogous? Therefore, we compare the realistic liquid bridge model with an equivalent simple linear irreversible contact model \cite{singh2013mesoscale} that would give the same macroscopic effect.  

The results in this paper are organized in three main parts. In Sec. \ref{sec:correlation} of this paper we study the effect of varying liquid bridge volume and surface tension of the liquid on the macroscopic properties, the focus being to find a micro-macro correlation from this study. Most strikingly, we see a well defined relationship between these micro parameters and the macro properties like the steady state cohesion of the bulk material and macro-torque required under shear, neglecting the effect of fluid depletion in shear bands \cite{mani2012fluid,mani2013liquid} in quasistatic flow. In Sec. \ref{sec:torque} of this paper we show the derivation of macro torque from the boundary shear stress. In this section we also compare this torque with the torque calculated from forces due to contacts on the wall particles. In Sec. \ref{sec:analogouslin} of this paper, we discuss about the analogy of two different contact models, with a goal to understand which parameters at microscopic scale would give the same macroscopic behavior of the system. 

\section{Model System}
\subsection{Geometry}
\begin{figure}[!h]
  \begin{center}
  {%
	\includegraphics[width=\columnwidth]{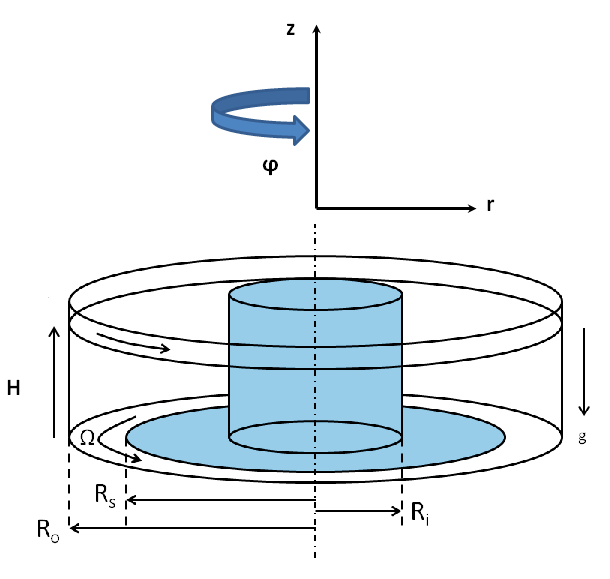}
   }%
  \end{center}
  \caption{Shear cell set-up.}
  \label{fig:setup}
\end{figure}

\textit{Split- Bottom Ring Shear Cell:} The set-up used for simulations consists of a shear cell with annular geometry and a split in the bottom plate, as shown in figure \ref{fig:setup}. Some of the earlier studies in similar rotating set-up include \cite{schollmann1999simulation,wang2012microdynamic,woldhuis2009wide}. The geometry of the system consists of an outer cylinder (radius $R_o$ = 110 mm) rotating around a fixed inner cylinder (radius $R_i$ = 14.7 mm) with a rotation frequency of $f_{rot}$ = 0.01 s$^{-1}$. The granular material is confined by gravity between the two concentric cylinders, the bottom plate, and a free top surface. The bottom plate is split at radius $R_s$ = 85 mm into a moving outer part and a static inner part. Due to the split at the bottom, a shear band is formed at the bottom. It moves inwards and widens as it goes up, due to the geometry. This set-up thus features a wide shear band away from the wall, free from boundary effects, since an intermediate filling height ($H$ = 40 mm) is chosen, so that the shear band does not reach the inner wall at the free surface. 
\par
In earlier studies \cite{singh2014effect,roy2015towards,singh2015role}, similar simulations were done using a quarter of the system (${0}^{\circ}$ $\leq$ $\phi$ $\leq$ ${90}^{\circ}$) using periodic boundary conditions. In order to save computation time, here we simulate only a smaller section of the system (${0}^{\circ}$ $\leq$ $\phi$ $\leq$ ${30}^{\circ}$) with appropriate periodic boundary conditions in the angular coordinate, unless specified otherwise. We have observed no noticeable effect on the macroscopic behavior in comparisons between simulations done with a smaller (${30}^{\circ}$) and a larger (${90}^{\circ}$) opening angle. Note that for very strong attractive forces, the above statement is not true anymore.

\subsection{Microscopic model parameters}\label{sec:var_par}
In presence of a small amount of liquid in a dense granular material, bridges are formed at the contact points between the particles. The surface energy of these bridges leads to an attractive force between the particles, which is absent in dry granular materials. Thus, wetting changes a granular system from one with only repulsive inter-particle interactions to one with both repulsive and attractive interactions \cite{scheel2008morphological}. With the change in microscopic physical interactions in wet granular materials, the macroscopic behavior is also expected to differ from the dry materials. Therefore, we choose to vary some of the characteristic specifications of a liquid bridge model to understand the effect on macroscopic properties. All  the particle specifications and the fixed interaction parameters for the contact models are given in table \ref{table}. All the variable interaction parameters which include the liquid bridge volume $V_b$ and the surface tension of the liquid $\gamma$ are discussed in this section.
\begin{table*}
\caption{Model parameters}
\label{table}       
\begin{tabular}{lll}
\hline\noalign{\smallskip}
Parameter & Symbol & Value  \\
\noalign{\smallskip}\hline\noalign{\smallskip}
Sliding friction coefficient & $\mu_p$ & 0.01 \\ 
Elastic stiffness & $k$ & 120 Nm$^{-1}$ \\  
Viscous damping coefficient & ${\gamma}_o$ & 0.5$\times $10$^{-3}$ kgs$^{-1}$ \\
Angular frequency  & $\omega$ &  0.01 s$^{-1}$ \\
Particle density  & $\rho$ &  2000 kgm$^{-3}$ \\
Mean particle diameter  & $d_p$ &  2.2 mm \\
Contact angle  & $\theta$ &  ${20}^\circ$\\
\noalign{\smallskip}\hline
\end{tabular}
\end{table*}

\subsubsection{Bulk saturation and liquid bridge volume}\label{sec:BulkSat}
The bulk material can be characterized by different states such as the dry bulk, adsorption layers, pendular state, funicular state, capillary state or suspension depending on the level of saturation \cite{mitarai2006wet,denoth1982pendular}. In this paper we intend to study the phenomenology of liquid bridge between particles in the pendular state, where the well separated liquid bridges exist between particle pairs without geometrical overlap. In this section, we discuss about the critical bulk saturation of granular materials and the corresponding liquid bridge volumes in the pendular state. 
\par
The bulk saturation $S^*$ is defined as the ratio of liquid volume to void volume of the bulk \cite{weigert1999calculation,pietsch1967fluÈssigkeitsvolumen,schubert1982kapillaritat}. The demarcation between the pendular state and the more saturated funicular state is given by the saturation $S^* \approx$ 0.3 \cite{weigert1999calculation}. For each particle pair with a liquid bridge, a dimensionless volume $\varphi^*$ can be defined as the ratio of the volume of the liquid bridge at the contact, $V_b$ to the volume of the two contacting particles, $2V_p$:
\begin{equation}
\varphi^* = \frac{V_{b}}{2V_p} = \frac{V_b}{2(\frac{\pi}{6}d_p^3)}
\end{equation}
Assuming the liquid is homogeneously distributed throughout the material, the bulk saturation $S^*$ is obtained from the dimensionless volume $\varphi^*$ and the bulk porosity $\epsilon$ from the following equation \cite{weigert1999calculation,pietsch1967fluÈssigkeitsvolumen,schubert1982kapillaritat}:
\begin{equation}
S^* = \pi\frac{1-\epsilon}{\epsilon^2}\varphi^* 
\end{equation}
With a bulk porosity of the material $\epsilon$ = 0.4 and a mean particle diameter $d_p$  of 2.20 mm, the maximum liquid bridge volume in the pendular regime is approximately 284 nl. In order to study the influence of liquid content on the macroscopic properties, we analyzed the system for the following set of liquid bridge volumes $V_b$:
\begin{equation}\label{vol_par}
V_b \ \in \ \left\{ {0, \ 1, \ 2,  \  4.2, \ 8, \ 14,  \ 20,  \   75,  \  140, \  200}\right\} \  \textnormal{nl},
\end{equation}
which are seen to be well within the pendular regime. We also calculate the liquid volume as a percentage of the total volume of the system ($V_t$) based on the number of contacts. The number of contacts represented as $C_L$ increases with increasing liquid bridge volume in the system and is measured approximately:
\begin{multline}\label{con_par}
C_L \ \in \ \left\{33010\right., \ 36214, \ 36855,  \  37585, \ 38306, \ 39101, \\
 \ 39511,  \   41526,  \  42595, \  \left.43328\right\} ,
\end{multline}
Therefore, the volume percentage of liquid in the system is given by ${\varphi}_b = \frac{{C_L}{V_b}}{V_t}$ and is approximately equal to:
\begin{multline}\label{vol_per}
{\varphi}_b \ \in \ \left\{0\right., \ 0.03, \ 0.07,  \  0.15, \ 0.29, \ 0.52, \\
  \ 0.75,  \   2.94,  \  5.63, \  \left.8.18\right\}. \ 
\end{multline}
In order to investigate the functional form of steady state cohesion beyond this state, a few more simulations for higher $V_b$ are done:
\begin{equation}\label{vol_par1}
V_b \ \in \ \left\{ {500, \ 1000}\right\} \  \textnormal{nl},
\end{equation}  
for which the pendular assumption is not valid anymore.
\subsubsection{Surface tension of liquid} 
Surface tension results from the greater attraction of liquid molecules towards each other than towards air. It is the tendency of liquids to lower their state of energy which makes it acquire the least possible surface area at the surface with higher inter liquid molecules attraction. As a result, cohesive properties of liquids are reflected in surface tension which makes it an interesting parameter to study. This effect will be discussed in detail in Sec. \ref{sec:contactmodel}. The effect of surface tension on the macroscopic properties is studied for the following range of surface tension values:
\begin{equation}\label{surf_par}
\gamma \ \in \  \left\{ {0,  \  0.020,  \ 0.040,  \   0.060}\right\} \  \textnormal{Nm}^{-1}.
\end{equation}
Surface tension of most of available liquid-air interfaces at 20$^{\circ}$C are in this range. To investigate the functional behavior of steady state cohesion beyond this state, a few more simulations for higher $\gamma$ are done:
\begin{equation}\label{surf_par1}
\gamma \ \in \  \left\{ {0.01,  \  0.10,  \  0.50,  \ 1.00}\right\} \  \textnormal{Nm}^{-1}.
\end{equation}
\subsection{Liquid bridge contact model}
\begin{figure}[!h]
  \begin{center}
  {%
	\includegraphics[width=\columnwidth]{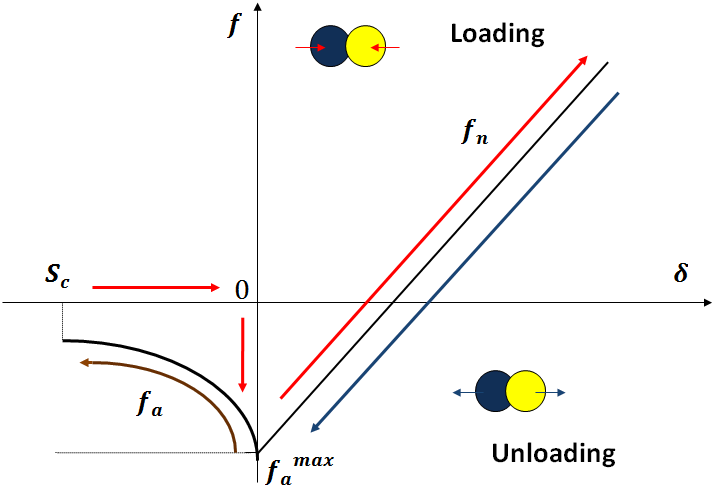}
   }%
  \end{center}
  \caption{Liquid capillary bridge model. The red lines represent the loading direction, the blue line represents the unloading direction when the particles are in contact and the brown line represents the unloading for the non-contact particles with short-range interaction force.}
  \label{fig:liquidbridge}
\end{figure}

The contact and non-contact forces for interacting particles can be described by a combination of an elastic contact model for the normal repulsive force and a non-linear irreversible adhesive non-contact model for the adhesive force. Figure \ref{fig:liquidbridge} represents a sketch of the combined liquid bridge contact model as a function of the overlap between the two particles.
 The liquid bridge adhesive force acts between the particles once the contact is established. When the particles are in contact, the attractive force is given by Eq. \eqref{max_force}. This is independent of the liquid bridge volume and depends on the surface tension of the liquid, radius of particles and contact angle. There is no cohesive force between the particles during approach. As the liquid bridge only forms once the particles come in contact with each other, the cohesive force starts acting and remains constant during overlap between particles $\delta >$ 0. Normal contact repulsive force acts between the particles in contact in addition, given by:
\begin{equation} \label{repulsive}
{f_n} = k\delta + {{\gamma}_o}\dot{\delta},
\end{equation}
where $k$ is the elastic stiffness, ${\gamma}_o$ is the viscous damping coefficient and $\delta$ is the overlap between the particles.
The normal contact forces for the liquid bridge model are explained in Sec. \ref{sec:contactmodel}
\subsubsection{Liquid bridge capillary force model} \label{sec:contactmodel}
The capillary pressure difference sustained across the liquid-air interface due to surface tension can be described by the non-linear Laplace-Young equation \cite{soulie2006influence}. This relates the pressure difference to the shape of the surface under the criterion of minimum Gibbs free energy \cite{erle1971liquid}. The capillary force in a pendular bridge originates from the axial component of this force. Another component that contributes to the capillary force is due to the hydrostatic pressure. Many previous studies have calculated capillary forces based on the numerical solution of the Laplace-Young equation and also reported experimental results \cite{willett2000capillary,soulie2006influence}. The magnitude of liquid bridge capillary force depends on the volume of the liquid bridge between the particles, the contact angle $\theta$, surface tension $\gamma$, the effective radius of the particles $r$ and the separation distance $S$, $S = -\delta$. With these parameters we approximate the inter-particle force $f_c$ of the capillary bridge according to \cite{willett2000capillary}. The experimental results are fitted by a polynomial to obtain the dependence of capillary forces on the scaled separation distance. During approach of the particles, the normal contact force for this model is given by:
\begin{align}\label{f_com1}
f = \begin{cases}
0 & \text{if $\delta < 0$}; \\
- {f_{a}}^{max} + f_n& \text{if $\delta \geq 0$}.\\
\end{cases}
\end{align}

During separation of the particles, the normal contact force for this model is given by:
\begin{align}\label{f_com2}
f = \begin{cases}
0 & \text{if $\delta < -S_c$}; \\
- f_a & \text{if $ -S_c \leq \delta < 0$};\\
- {f_{a}}^{max} + f_n & \text{if $\delta \geq 0$},
\end{cases}
\end{align}

where $f_n$ is the normal repulsive force given by Eq. \eqref{repulsive}. The adhesive force for the liquid bridge model is the capillary force given by:
\begin{equation} \label{f_cap1}
{f_a} = ({f_a})_{liq} = \frac{{{({f_a}^{max})}_{liq}}(\frac{2r}{d_p})}{1+1.05\bar{S}+ 2.5\bar{S}^2},
\end{equation}
where the separation distance is normalised as $\bar{S} = S\sqrt{({r}/{V_b})}$, $S$ being the separation distance. The maximum capillary force between the particles when they are in contact ($S$ = 0) is given by:
\begin{equation} \label{max_force}
{({f_a}^{max})}_{liq} = \pi{d_p}{\gamma}{\cos\theta},
\end{equation}
where, $d_p$ is the mean particle diameter.
The effective radius of two interacting spherical particles of different size can be estimated as the harmonic mean of the two particle radii according to the Derjaguin approximation \cite{deryaguin1934untersuchungen}, yielding the effective radius:
\begin{equation}
r = \frac{2{r_i}{r_j}}{r_i+r_j},
\end{equation}
however, the mean size is not varied here.
This model equation is applicable for mono-disperse particles \cite{schwarze2013rheology,willett2000capillary} which has been actually extended to poly-disperse system of particles Ref. \cite{herminghaus2005dynamics}. As proposed by \cite{lian1993theoretical}, the critical separation distance $S_c$ between the particles before the bridge ruptures is given by:
\begin{equation} \label{rup_dist}
(S_c)_{liq} = \bigg{(}1 + \frac{\theta}{2}\bigg{)}V_b^{1/3}
\end{equation}
\begin{figure}[!h]
  \begin{center}
  {%
	\includegraphics[width=\columnwidth]{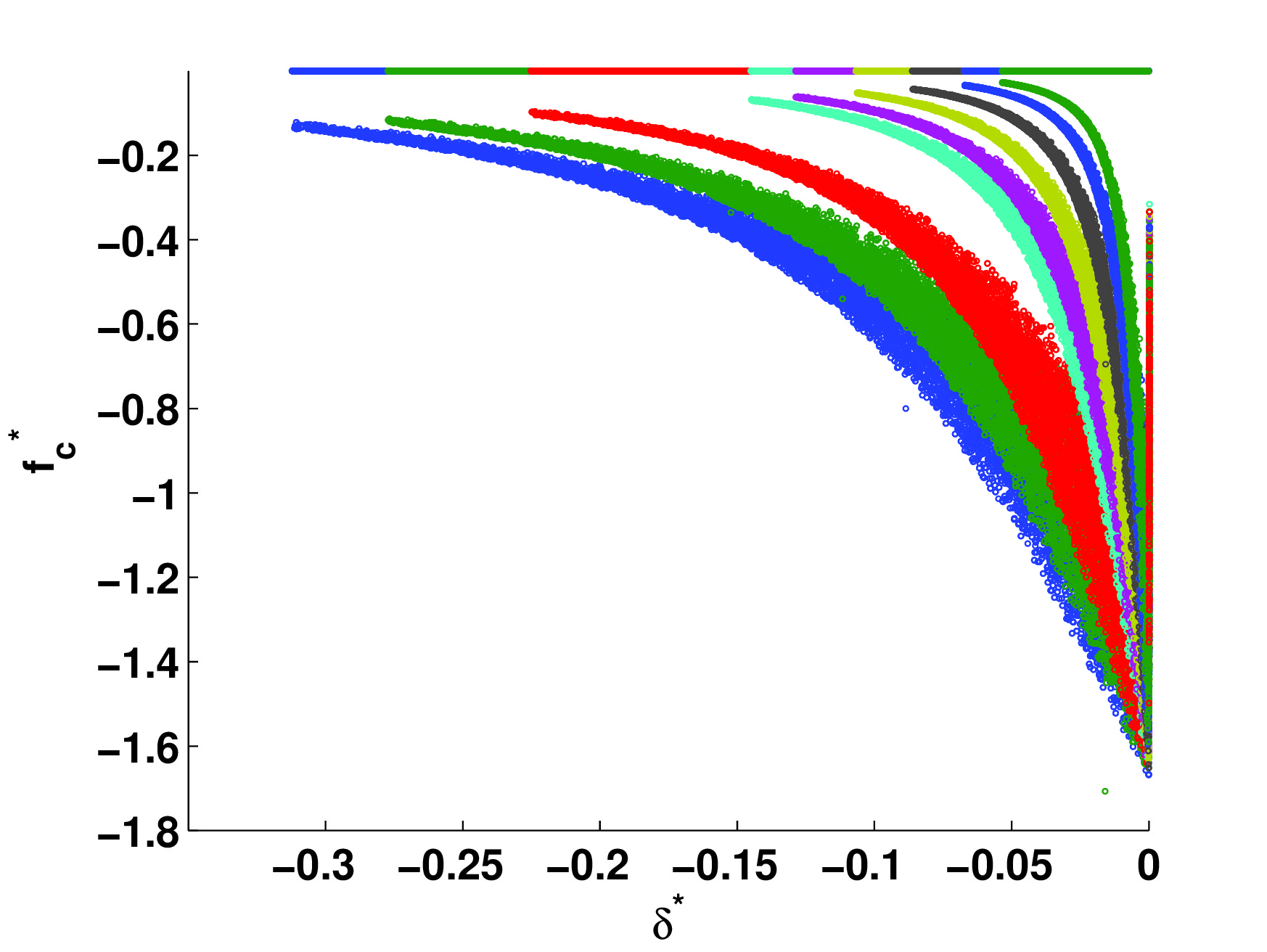}
   }%
  \end{center}
  \caption{${f_c}^*$ as a function of ${\delta}^*$. Different colors represent different liquid bridge volumes.}
  \label{fig:liqsim}
\end{figure}
The liquid bridge capillary force as a function of separation distance is shown in figure \ref{fig:liqsim} for different liquid bridge volumes. The liquid bridge capillary force decreases in magnitude with increasing separation distance between the particles till the bridge ruptures. The rupture distance is proportional to $V_b^{1/3}$ as stated in Eq. \eqref{rup_dist}. 
\subsubsection{Linear irreversible contact model} \label{sec:contactmodel1}
\begin{figure}[!h]
  \begin{center}
  {%
	\includegraphics[width=\columnwidth]{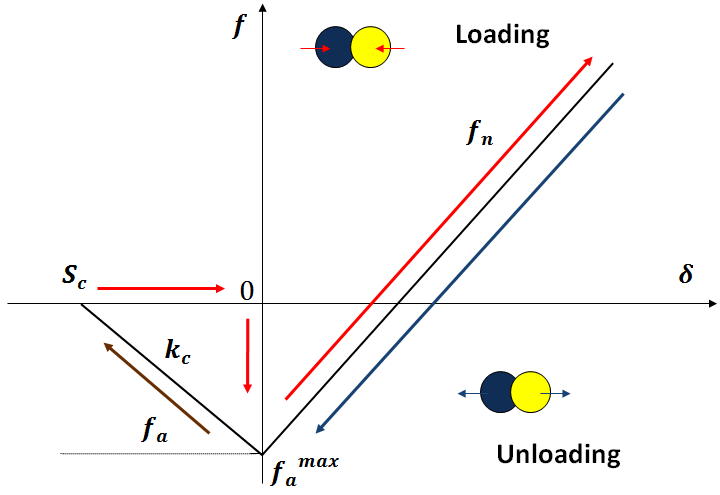}
   }%
  \end{center}
  \caption{Linear irreversible contact model. The red lines represent the loading direction, the blue line represents the unloading direction when the particles are in contact and the brown line represents the unloading for the non-contact particles with short-range interaction force.}
  \label{fig:linearmodel}
\end{figure}
In Sec. \ref{sec:analogouslin} we introduce a simple linear irreversible contact model as proposed by \cite{singh2013mesoscale} and shown in figure \ref{fig:linearmodel} which can be compared with the non-linear liquid bridge interaction model. For the linear irreversible contact model, the normal forces between particles during approach and separation are given by Eqs. \eqref{f_com1} and \eqref{f_com2} respectively, where for the linear irreversible contact model,
\begin{equation} \label{f_lin1}
{f_a} = ({f_a})_{lin} = {({f_a}^{max})}_{lin} + k_c{\delta},
\end{equation}
\begin{equation} \label{f_lin2}
(S_c)_{lin} = {({f_a}^{max})}_{lin}/k_c,
\end{equation}

where ${({f_a}^{max})}_{lin}$ is the maximum adhesive force and $k_c$ is the adhesive stiffness. The tangential force contact model is explained in details in our earlier studies \cite{roy2015towards}. 

\subsection{Dimensional analysis}
To formulate all the modeling equations in a constructive way, we express them in nondimensionalized form. All the length scale parameters are scaled by the mean particle diameter $d_p$ = 2.20 mm. The forces are scaled in terms of the gravitational force acting on a single particle $f_g$ = $V_p{\rho}g$ $\approx$  1.0939 $\times$ 10$^{-4}$ N. Table \ref{table1} shows all the parameters in their dimensionless form and the corresponding scaling terms used in the equations. The angular rotation of the shear cell after a given time to study the dynamic evolution of torque is scaled in terms of radians covered in one complete rotation ($2\pi$). The dynamics of the system can be characterized by the time scale defined by the contact duration between two particles $t_c = \sqrt{\frac{m_p}{k}}$, where ${m_p}$ is the mean mass of a particle. Since we do all our macro-rheology analysis in steady state, characterization of dynamics of the system is not required. The main objectives of nondimensionalization is to simplify the equations in terms of unit less quantities and define the system intrinsically.

\begin{table*}
\caption{Non-dimensionalization of parameters}
\label{table1}       
\begin{tabular}{llll}
\hline\noalign{\smallskip}
Parameter & Symbol & Scaled term & Scaling term  \\
\noalign{\smallskip}\hline\noalign{\smallskip}
Capillary force & $f_c$ & ${f_c}^*$ & $f_g$ \\ 
Particle overlap & $\delta$ & ${\delta}^*$ & $d_p$ \\ 
Shear stress & $\tau$ & ${\tau}^*$ & $f_g/{d_p}^2$ \\
Pressure & $P$ & ${P}^*$ & $f_g/{d_p}^2$ \\ 
Steady state cohesion & $c$ & ${c}^*$ & $f_g/{d_p}^2$ \\ 
Liquid bridge volume & ${V_b}$ & ${V_b}^*$ & ${d _p}^3$ \\ 
Surface tension & $\gamma$ & ${\gamma}^*$ & $f_g/{d_p}$ \\ 
Rupture distance & $S_c$ & ${S_c}^*$ & $d_p$ \\ 
Torque & $T_z$ & ${T_z}^*$ & ${f_g}{d_p}$ \\ 
Angular rotation & ${\theta_{rot}}$ & ${\theta_{rot}}^*$ & $2{\pi}$ \\ 
Adhesive Energy & $E$ & ${E}^*$ & ${f_g}{d_p}$ \\ 
\noalign{\smallskip}\hline
\end{tabular}
\end{table*}
\section{Micro macro transition}
To extract the macroscopic properties, we use the spatial coarse-graining approach detailed in \cite{luding2008effect,luding2008constitutive,luding2011critical}. The averaging is performed over toroidal volume, over many snapshots of time assuming rotational invariance in the tangential $\phi$-direction. The averaging procedure for a three dimensional system is explained in \cite{luding2008constitutive,luding2011critical}. This spatial coarse-graining method was used earlier in \cite{singh2014effect,singh2013mesoscale,roy2015towards,singh2015role,luding2011critical}. The simulation is run for 200 s and temporal averaging is done when the flow is in steady state, between 80 s to 200 s, thereby disregarding the transient behavior at the onset of the shear. 

\subsection{Steady state cohesion and its correlation with liquid bridge volume and surface tension}
\label{sec:correlation}
\begin{figure}[!h]
  \begin{center}
  {%
	\includegraphics[width=\columnwidth]{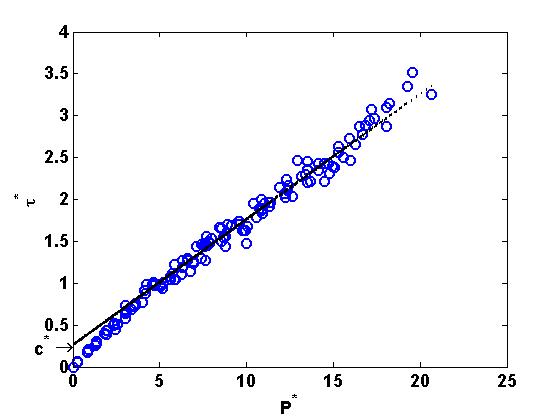}
   }%
  \end{center}
\caption{Shear stress ${\tau}^*$ plotted against pressure ${P}^*$. The dotted line represents the fitting function as given by Eq. \eqref{shear_stress} for ${P}^*$ $>$ 4.42 Pa where $\mu$ = 0.15 is the macroscopic friction coefficient, $c^*$ = 0.2655 for $V_b$ = 75 nl and $\gamma$ = 0.020 Nm$^{-1}$.}
  \label{fig:shearstr}
\end{figure}

\begin{figure}[!h]
  \begin{center}
  {%
	\includegraphics[width=\columnwidth]{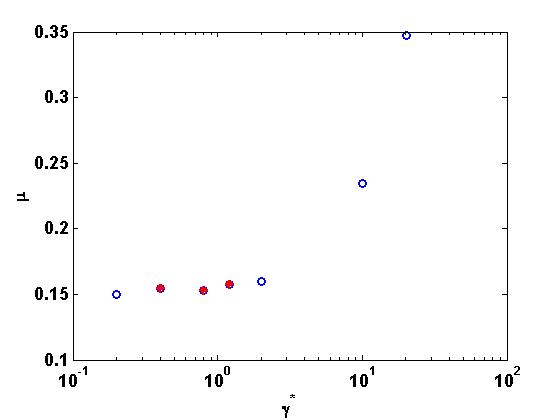}
   }%
  \end{center}
  \caption{Macroscopic friction coefficient $\mu$ as a function of ${\gamma}^*$ for $V_b$ = 75 nl. The solid symbols represent the range of surface tension for our simulations below.}
   \label{fig:mumacrosurf}
\end{figure}

\begin{figure}[!h]
  \begin{center}
  {%
	\includegraphics[width=\columnwidth]{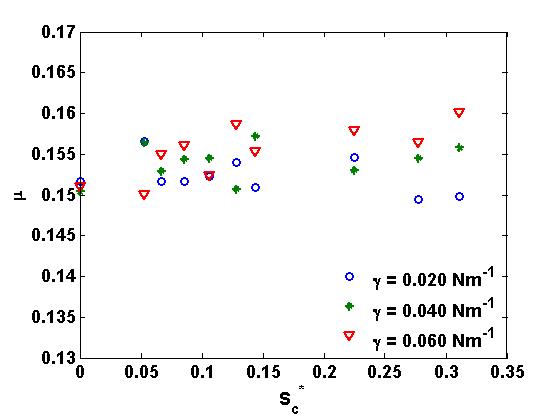}
   }%
  \end{center}
  \caption{Macroscopic friction coefficient $\mu$ as a function of ${S_c}^*$ for $\gamma$ = 0.020 Nm$^{-1}$. }
  \label{fig:mumacrorup}
\end{figure}

In earlier studies \cite{schwarze2013rheology,roy2015towards,luding2008effect,luding2008constitutive}, the shear band region was identified by the criterion of large strain rate, \textit{e.g.} higher than a critical strain rate of 0.08 s$^{-1}$. In this paper, the shear band center region is defined by strain rates higher 80$\%$ of the maximum for different heights in the shear cell. Figure \ref{fig:shearstr} displays the dependence of scaled yield stress ${\tau}^*$ for the particles in the shear band region on scaled pressure $P^*$ for 75 nl liquid bridge volume. A linear trend is observed neglecting the different behavior for data at very low pressure ($P^*<4.42$). This is fitted well by a linear function:
\begin{equation}\label{shear_stress}
{\tau}^* = {\mu}{P}^* + {c}^*
\end{equation}
where $\mu$ is the macroscopic friction coefficient and $c^*$ is the steady state cohesion obtained from the plot. Next, we fit the data for shear stress as a function of pressure as given by Eq. \eqref{shear_stress} and obtain the value of steady state cohesion and macroscopic friction $\mu$. The macroscopic friction coefficient is constant for lower surface tension, including ${\gamma}^* = 0$ for linear elastic model (not shown in figure), but increases for ${\gamma}^* \gtrsim 2$ for a given liquid bridge volume as shown in figure \ref{fig:mumacrosurf}. When the surface tension of the material is very high (${\gamma}^* \gtrsim 1.00$), materials protrude out of the top surface to form a hump in the region of the shear band (data not shown). For our analysis of surface tension in the range 0.020-0.040 Nm$^{-1}$, the macroscopic friction coefficient is constant at $\mu \simeq$ 0.15. In this range, the macroscopic friction coefficient is also independent of the liquid bridge volume as shown in figure \ref{fig:mumacrorup}.


\par
For dry cohesionless systems, the dependence of shear stress on pressure is linear without an offset, \textit{i.e.} $c^*$ = 0. In the presence of interstitial liquid between the particles in the pendular regime, cohesive forces increase with increasing liquid bridge volume. This results in a positive steady state cohesion $c^*$ as given by Eq. \eqref{shear_stress}, see figure \ref{fig:shearstr}. 

\begin{figure}[!h]
	\includegraphics[width=\columnwidth]{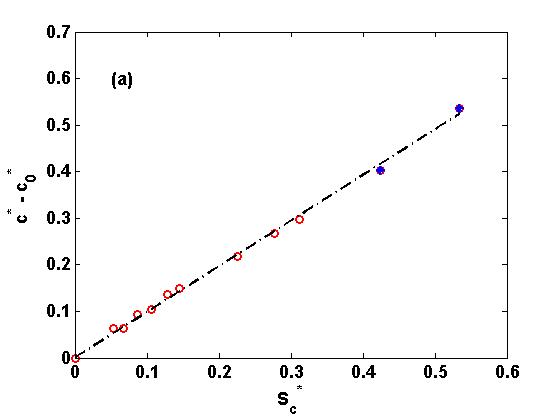}
	\includegraphics[width=\columnwidth]{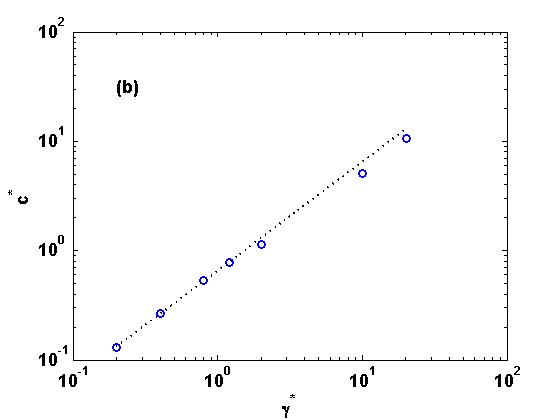}
\caption{(a) ${c^*-{c_0}^*}$ as a function of ${S_c}^*$ for $\gamma = 0.020$ Nm$^{-1}$. The dotted line represents the fitting function given by Eq. \eqref{tot_coh1}. The data with solid symbols represent the liquid bridge volume outside the pendular regime. (b) $c^*$ as a function of ${\gamma}^*$ for $V_b$ = 75 nl. The dotted line represents the fitting function given by Eq. \eqref{c_surf}.}
\label{fig:macrocoh}
\end{figure}
Earlier studies on wet granular materials have shown that the presence of liquid bridges between the particles results in an increasing steady state cohesion of the materials \cite{schwarze2013rheology,gladkyy2014comparison,richefeu2007shear,roy2015towards}. Our earlier studies show that the steady state cohesion $c^*$ increases non-linearly with increasing liquid bridge volume. Here, the steady state cohesion is studied in more detail, including very small liquid bridge volumes, including the (practically impossible) limit of 0 nl liquid bridge volume as given in Eq. \eqref{vol_par}. Note that there is a finite cohesive strength for $V_b \to$  0 nl liquid bridge volume. This is due to the microscopic capillary bridge force that acts between particles even at 0 nl liquid bridge volume as given by Eq. \eqref{max_force}. This is called the steady state critical cohesion ${c_0}^*$ for a given surface tension of liquid. This value depends on the maximum force acting between two particles when they are in contact as given by Eq. \eqref{max_force}. The additional cohesion for higher liquid bridge volume is due to the non-contact capillary forces between the particles that are active upto the distance when the liquid bridge ruptures. This is dependent on the surface tension of the liquid and the volume of the liquid bridge. Thus, the steady state cohesion of granular materials for a given liquid bridge volume can be written as:
\begin{equation}\label{tot_coh}
c^* = {c_0}^* + {c'}^*
\end{equation}
 where ${{c}'}^*$ is the additional cohesion for liquid bridge volume $V_b > 0$. Figure \ref{fig:macrocoh}(a) shows $({c}^*-{c_0}^*)$ as a linear function of ${{S_c}^*}$, fitted by:
\begin{equation}\label{tot_coh1}
{c}^* - {c_0}^* = a{S_c}^* 
\end{equation}
where $a$ = 0.9805 for $\gamma = 0.020$ Nm$^{-1}$. In the next section we study the dependence of this constant on the surface tension of liquid.
\par
Figure \ref{fig:macrocoh}(b) shows the dependence of steady state cohesion on ${\gamma}^*$ for $V_b =$ 75 nl. The steady state cohesion can be described by:
\begin{equation}\label{c_surf}
\ln{c}^* = \alpha\ln{\gamma}^* + k
\end{equation}
where $\alpha \approx 1.00$, $k = - 0.4240$. Therefore, the steady state cohesion is linearly proportional to the surface tension and can be written as:
\begin{equation}\label{c_surf1}
{c}^* = b{{\gamma}^*}
\end{equation}
where $b = \exp{(k)}$. The above equation is valid in the limit of zero surface tension (${\gamma}^* = 0$) which represents the simple linear elastic contact model. For higher surface tension of liquid, the results deviate from the fitted function of linear dependence as seen from figure \ref{fig:macrocoh}(b). As given by Eq. \eqref{tot_coh1} and \eqref{c_surf}, the steady state cohesion is dependent on liquid bridge volume expressed in terms of  maximum interaction distance ${S_c}^*$ between the particles and the maximum adhesive force expressed in terms of surface tension of the liquid ${\gamma}^*$. So in the later sections of this paper we study the dependence of macroscopic parameters on the micro parameters ${S_c}^*$ representing scaled rupture distance and ${\gamma}^*$ representing scaled maximum force for all contact models.

\begin{figure}[!h]
  \begin{center}
  {%
	\includegraphics[width=\columnwidth]{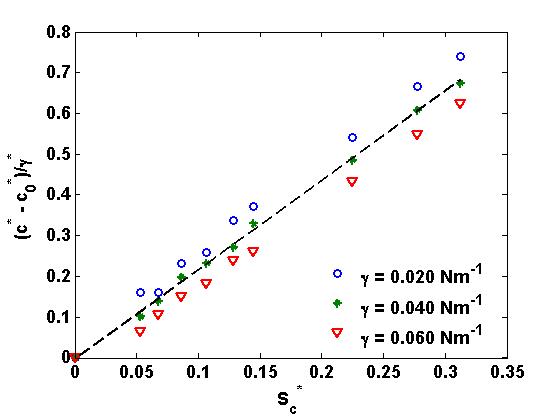}
   }%
  \end{center}
  \caption{$\frac{{c}^* - {c_0}^*}{{{\gamma}^*}}$ as a function of ${S_c}^*$ for different surface tension of liquid. The dotted line represents the fitting function given by Eq. \eqref{eq:c_rupture}.}
  \label{fig:c_rupture}
\end{figure}

Figure \ref{fig:c_rupture} shows the dependence of $\frac{{c}^* - {c_0}^*}{{{\gamma}^*}}$ on ${S_c}^*$ for different surface tension of liquid. The scaled steady state cohesion is a linearly dependent on the rupture distance as shown in the figure. This can be fitted by a straight line equation given by:
 \begin{equation}
\frac{{c}^* - {c_0}^*}{{{\gamma}^*}} = \frac{{c}^* - {c_0}^*}{{{{({f_a}^{max})}_{liq}}^*}/({\pi}cos{\theta})} = p{S_c}^* 
\label{eq:c_rupture}
\end{equation}
where $p$ = 2.1977 as obtained from the fitting shown in figure \ref{fig:c_rupture}; the offset is very small and can be neglected.  
This subsection shows that the macroscopic characteristics of the liquid bridge model are determined by the maximum interacting force between the particles and the rupture distance. The steady state cohesion scales linearly with the surface tension of liquid \textit{i.e.} the maximum force between the particles. For a given maximum force, the cohesion scaled with the surface tension of liquid is also a linear function of the rupture distance of the liquid bridge.
\subsection{Macroscopic torque analysis from the microscopic parameters}
\label{sec:torque}
The strength, cohesion and flow properties of granular materials are strongly influenced by the presence of capillary cohesion. Due to the cohesive properties of these wet materials, the shear stress increases and, as a result, partially saturated wet materials require higher torques for deformation (shear) \textit{e.g.} in a shear cell. Loosely speaking, torque is a measure of the shear stress or force acting on the particles at the wall and thus can be used to find an estimate of shear stress in the shear band. To study solely the effect of capillary cohesion on the torque, the other parameters like the particle friction is kept very small in our simulations, with $\mu_{p}$ = 0.01. Earlier studies \cite{gladkyy2014comparison,roy2015towards,bouwman2005relation,obraniak2012model} show that the average torque acting on the rotating part of the shear cell increases with increasing moisture content. In this section we perform a detailed analysis of the macroscopic torque as a function of the micro parameters in order to understand its connection with the steady state cohesion of the material. 
\par

The walls and the bottom plates of the shear cell consist of particles with a prescribed position. The particles forming the inner wall are stationary while the particles forming the outer wall rotate around the z-axis with frequency $f_{rot}$. All the particles forming the inner and outer wall are identified as $\mathcal{C}_\mathrm{inner}$ and $\mathcal{C}_\mathrm{outer}$, respectively. The macroscopic torque is calculated based on the contact forces on the fixed particles on the moving (outer) and stationary (inner) parts of the shear cell. Thus the net inner and outer torque are calculated by summing up the torques for all the contacts with respect to the axis of rotation of the shear cell. The net torque is obtained from the difference between the outer wall torque and the inner wall torque. We multiply the total torque by a factor of $\frac{2\pi}{\pi/6}$ in order to get the torque for the whole system from the obtained torque of our simulations in a 30$^\circ$ section. Thus the torque is given by:
\begin{multline} \label{torque}
\vec{{T}} = \frac{2\pi}{\pi/6}\big[{\big(\mathop{\sum_{i=1}^{N}\sum\nolimits_{j \in \mathcal{C}_\mathrm{outer}} {{\vec{c}_{i,j}}}\times {\vec{f}_{i,j}}}}\big) -  \\
\big({\mathop{\sum_{i=1}^{N}\sum\nolimits_{j \in \mathcal{C}_\mathrm{inner}} {{\vec{c}_{i,j}}}\times {\vec{f}_{i,j}}}}\big)\big] ,
\end{multline}
where $N$ represents the number of particles, $\vec{c}_{ij}$ is the position of the contact point and $\vec{f}_{ij}$ is the interaction force. Only the $z$-component of the torque vector ($T_z$) is of interest as required for shearing the cell in angular direction. 

\begin{figure}[!h]
  \begin{center}
  {%
	\includegraphics[width=\columnwidth]{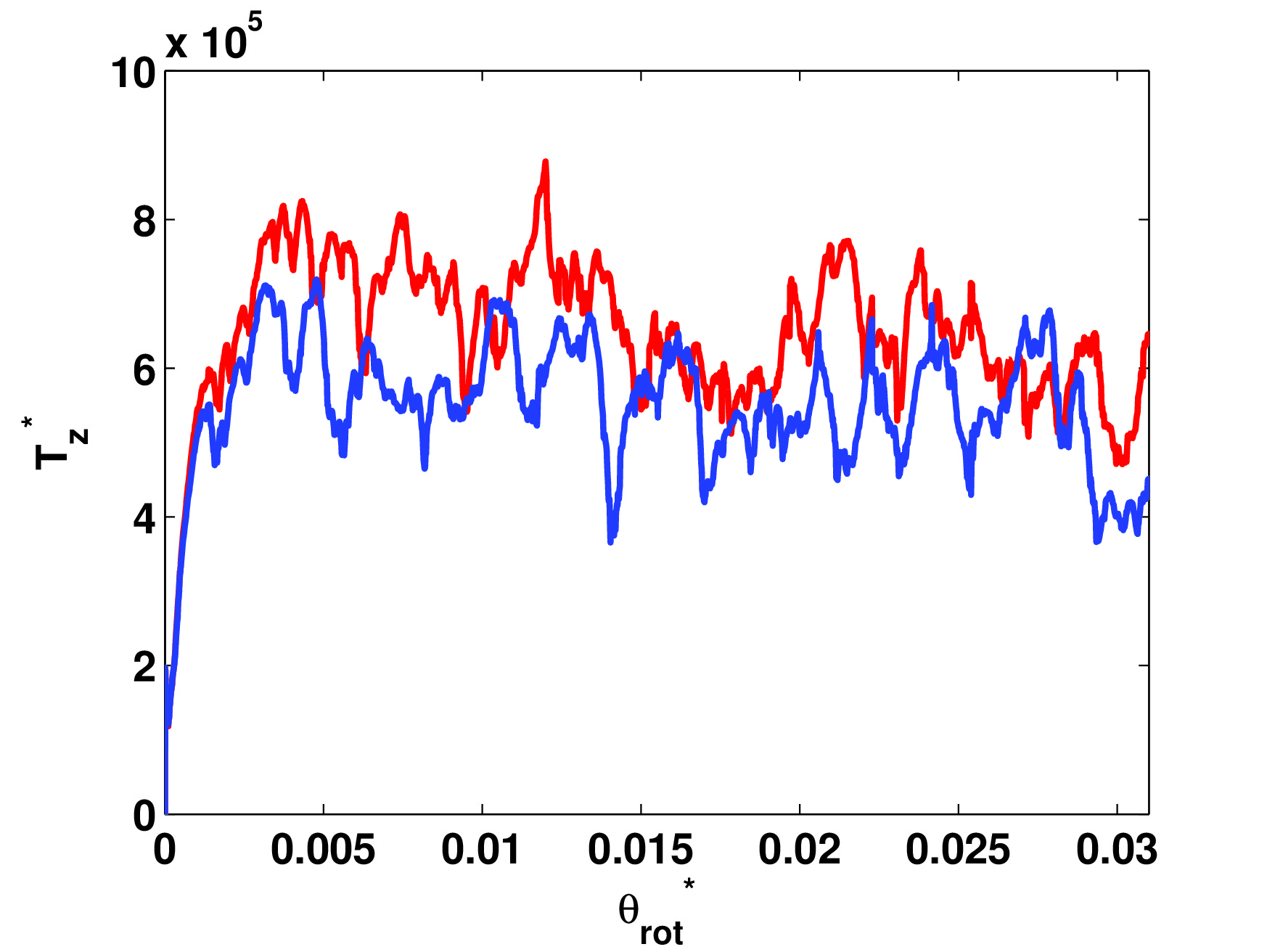}
   }%
  \end{center}
  \caption{${T_z}^*$ as a function of scaled angular rotation ${\theta_{rot}}^*$ for surface tension of liquid $\gamma =$ 0.020 Nm$^{-1}$ for $V_b =$  4.2 nl (blue) and $V_b =$  200 nl (red).}
  \label{fig:Torque_angle}
\end{figure}

We compare our results with the experimental results as given by \cite{wortel2014rheology} from the evolution of torque as a function of the angular rotation as shown in figure \ref{fig:Torque_angle}. This is in good agreement with the magnitude and angular rotation required for steady state torque evolution as given in \cite{wortel2014rheology}, considering the different rotation rate and different friction in the systems. 
\par
\begin{figure}[!h]
  \begin{center}
  {%
	\includegraphics[width=\columnwidth]{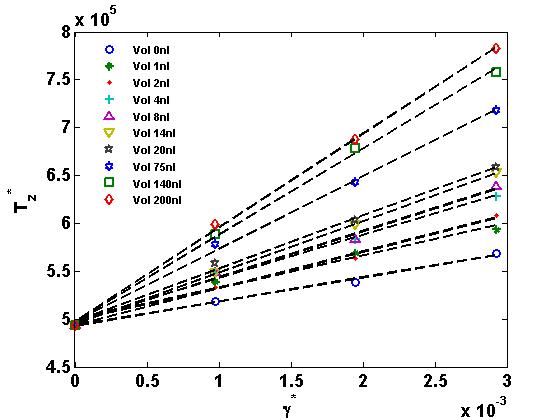}
   }%
  \end{center}
  \caption{${T_z}^*$ as a function of ${\gamma^*}$. The dotted lines represent the fitting functions for different liquid bridge volumes given by equation ${T_z}^* = l{{\gamma^*}} + t$ where $t$ = 4.964 $\times 10^5$  and $l$ increases with increasing liquid bridge volume.}
  \label{fig:Tor_Surf}
\end{figure}

Figure \ref{fig:Tor_Surf} shows ${{T}_z}^*$ as a function of ${{\gamma}^*}$ for different liquid bridge volumes. We observe that the resultant torque depends linearly in the surface tension of the liquid. The fit parameter $l$ from the figure, the rate of increase of torque with surface tension, depends on the liquid bridge volume.
\par
Next, we compare the results of the steady state cohesion as obtained from the fitting function explained in Sec. \ref{sec:correlation} with the calculated (measured) torque. We write the scalar form of the torque on the wall derived from steady state cohesion as ${{T}_z}^{macro}$:
\begin{equation}\label{T_corell}
{{T}_z}^{macro} = \left[\int_{A_o}{r\,dA} - \int_{A_i}{r\,dA}\right]{({\mu}P_{avg} + c)},
\end{equation}
where $A_o$ denotes the outer wall surface, $A_i$ denotes the inner wall surface and $P_{avg}$ is the mean pressure inside the shear band approximately 250 Pa for a filling height of 39 mm. Eq. \eqref{T_corell} can be simplified to the form:
\begin{equation}\label{T_corell1}
{{T}_z}^{macro} = M(\mu{P_{avg}} +  c)
\end{equation}
where $M = \big{[}2{\pi}H({R_o}^2 - {R_i}^2)  +  \frac{2}{3}{\pi}({R_o}^3 + {R_i}^3 - 2{R_s}^3)\big{]}\approx$ 0.0031\,$\mathrm{m}^3$ for the given geometry is equal to fitting parameter $t/({{\mu}{P_{avg}}})$, $t$ is the fit parameter, see figure \ref{fig:Tor_Surf}. Assuming ${{T}_z} = {{T}_z}^{macro}$, an equivalent steady state cohesion as obtained from the calculated torque can be given as:
\begin{equation}\label{c_eq}
c_{eq} = {{T}_z}/M -  \mu{P_{avg}}
\end{equation}

\begin{figure}[!h]
  \begin{center}
  {%
	\includegraphics[width=\columnwidth]{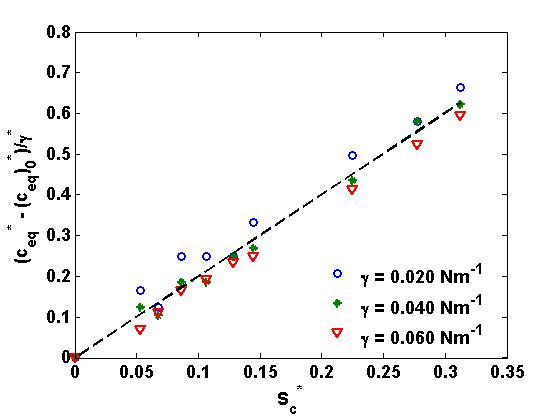}
   }%
  \end{center}
  \caption{$\frac{{c_{eq}}^* - {(c_{eq})_0}^*}{{{\gamma}^*}}$ as a function of ${S_c}^*$ for different surface tension of liquid where $c_{eq}$ is given by Eq. \eqref{c_eq}. The dotted line represents the fitting function as given by Eq. \eqref{eq:Torque_rupture1}.}
  \label{fig:Torque_Rup}
\end{figure}

\begin{figure}[!h]
  \begin{center}
  {%
	\includegraphics[width=\columnwidth]{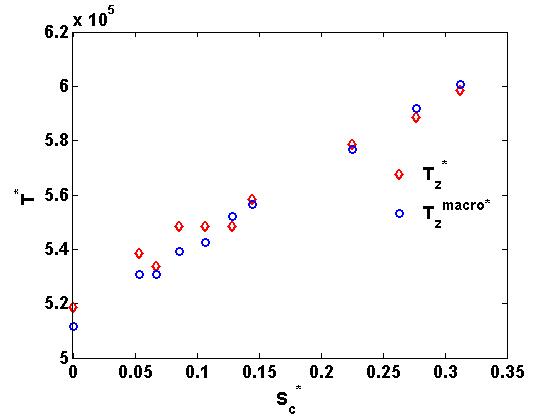}
   }%
  \end{center}
  \caption{Torque calculated numerically scaled as ${{T}_z}^*$ as compared with the scalar form of scaled macro torque ${{{T}_z}^{macro}}^*$ as calculated from the wall shear stress as given by Eq. \eqref{T_corell1}.}
  \label{fig:Tor_compare}
\end{figure}

Figure \ref{fig:Torque_Rup} shows the dependence of the non-di\-men\-sio\-na\-lised value $\frac{{c_{eq}}^* - {(c_{eq})_0}^*}{{{\gamma}^*}}$ on ${S_c}^*$ for different surface tension. ${(c_{eq})_0}^*$ is the equivalent steady state cohesion as obtained from Eq. \eqref{c_eq} for the torque of a 0 nl liquid bridge. This can be fitted by a straight line:
\begin{equation}
\frac{{c_{eq}}^* - {(c_{eq})_0}^*}{{{\gamma}^*}} = e{S_c}^*, 
\label{eq:Torque_rupture1}
\end{equation}
where $e$ = 2.0062 is a fit parameter, see figure \ref{fig:Torque_Rup}, and the offset is very small and can be neglected. Eq. \eqref{eq:Torque_rupture1} shows equivalent steady state cohesion as obtained from the torque is also linearly dependent on ${S_c}^*$. The fitting parameter $e$ of this equation shows a close similarity with the fitting parameter $p$ of Eq. \eqref{eq:c_rupture}. Alternatively, figure \ref{fig:Tor_compare} shows a comparison of the two torque given by the scalar $z$-component of Eq. \eqref{torque} and Eq. \eqref{T_corell1} for surface tension of liquid 0.020 Nm$^{-1}$. These results show that the steady state cohesion and torque are related by Eq. \eqref{T_corell1}.
\par
In conclusion, this subsection shows that the measured torque can be translated to the local steady state macro-rheology parameters via a simple factor $M$ (a measure of the resultant arm-length times surface area) which depends only on the geometry of the system.


\section{An analogous linear irreversible contact model for cohesive particles}
\label{sec:analogouslin}
\begin{figure}[!t]
  \begin{center}
  {%
	\includegraphics[trim = 0 0 0 0, clip,width=\columnwidth]{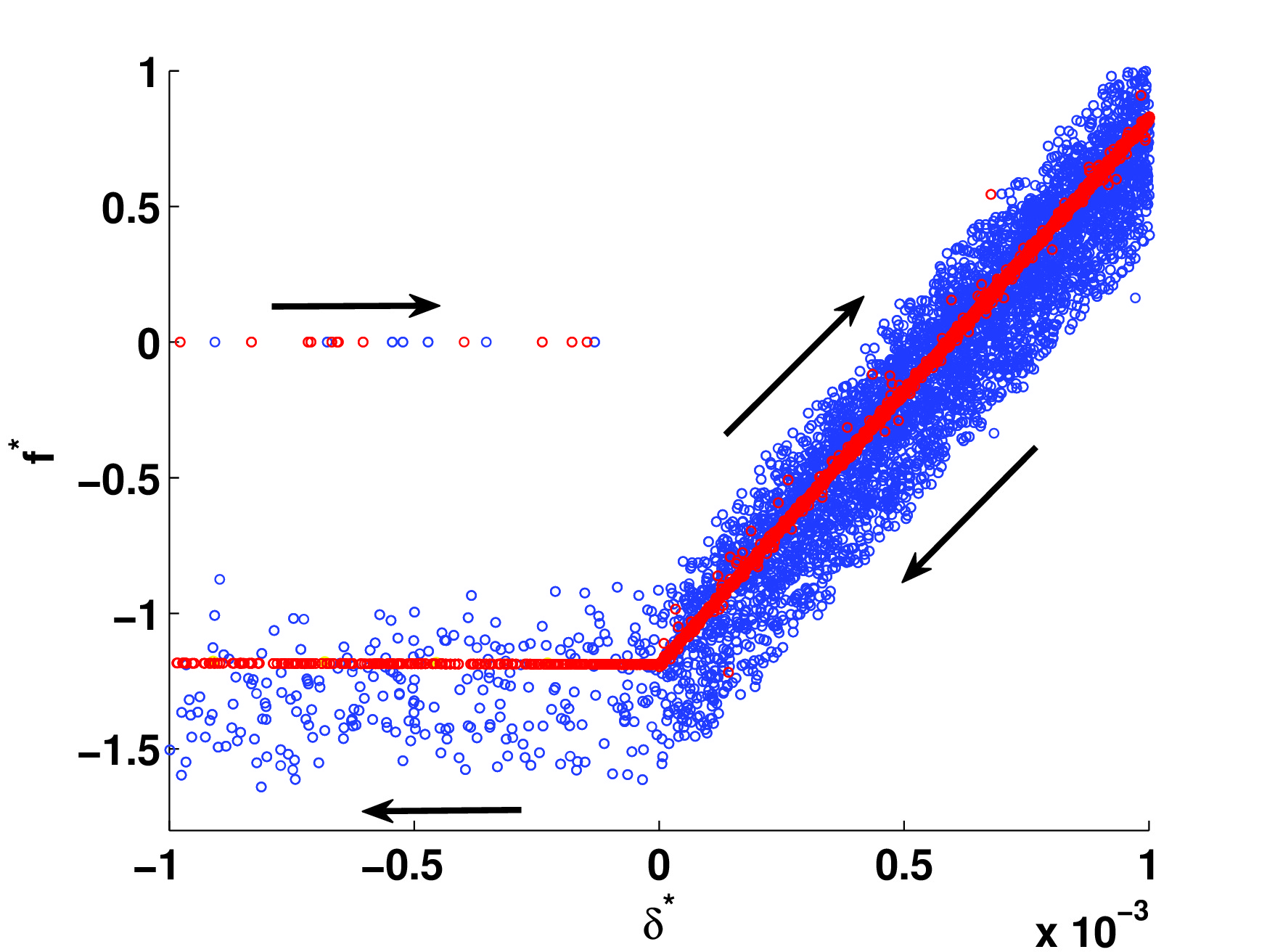}
   }%
  \end{center}
  \caption{Force-overlap diagram for the liquid bridge model (blue) as compared with the linear-irreversible contact model (red). The arrow shows the loading and the unloading directions for all forces. The schematic diagram for the same are given in figures \ref{fig:liquidbridge} and \ref{fig:linearmodel} respectively.}
  \label{fig:analogous}
\end{figure}

In this section we aim to determine the key microscopic parameters for a linear irreversible contact model \cite{singh2013mesoscale} that is macroscopically analogous to the liquid bridge contact model used before. An explanation of the linear irreversible contact model is given in \cite{singh2013mesoscale}. Unlike the liquid bridge contact model, the force for the linear irreversible contact model is simple and faster to compute. Figure \ref{fig:analogous} shows the force-overlap distribution for the two contact models showing the loading and unloading directions of forces which are reversible at ${\delta}^* > 0$ and irreversible at ${\delta}^* < 0$.
\par
As discussed in Sec. \ref{sec:correlation}, the steady state cohesion for the liquid bridge model is controlled by the rupture distance of the liquid bridge, which is proportional to the liquid bridge volume, and the magnitude of the maximum interaction force, which is governed by the surface tension of the liquid. Assuming that the non-linear liquid bridge capillary force can be replaced by a simple irreversible linear adhesive force between the particles with the same  macro characteristics, we compare the steady state cohesion of the two models in Sec. \ref{sec:analogous}.
\subsection{Equal maximum force and interaction distance}\label{sec:analogous}
\begin{figure}[!t]
  \begin{center}
  {%
	\includegraphics[width=\columnwidth]{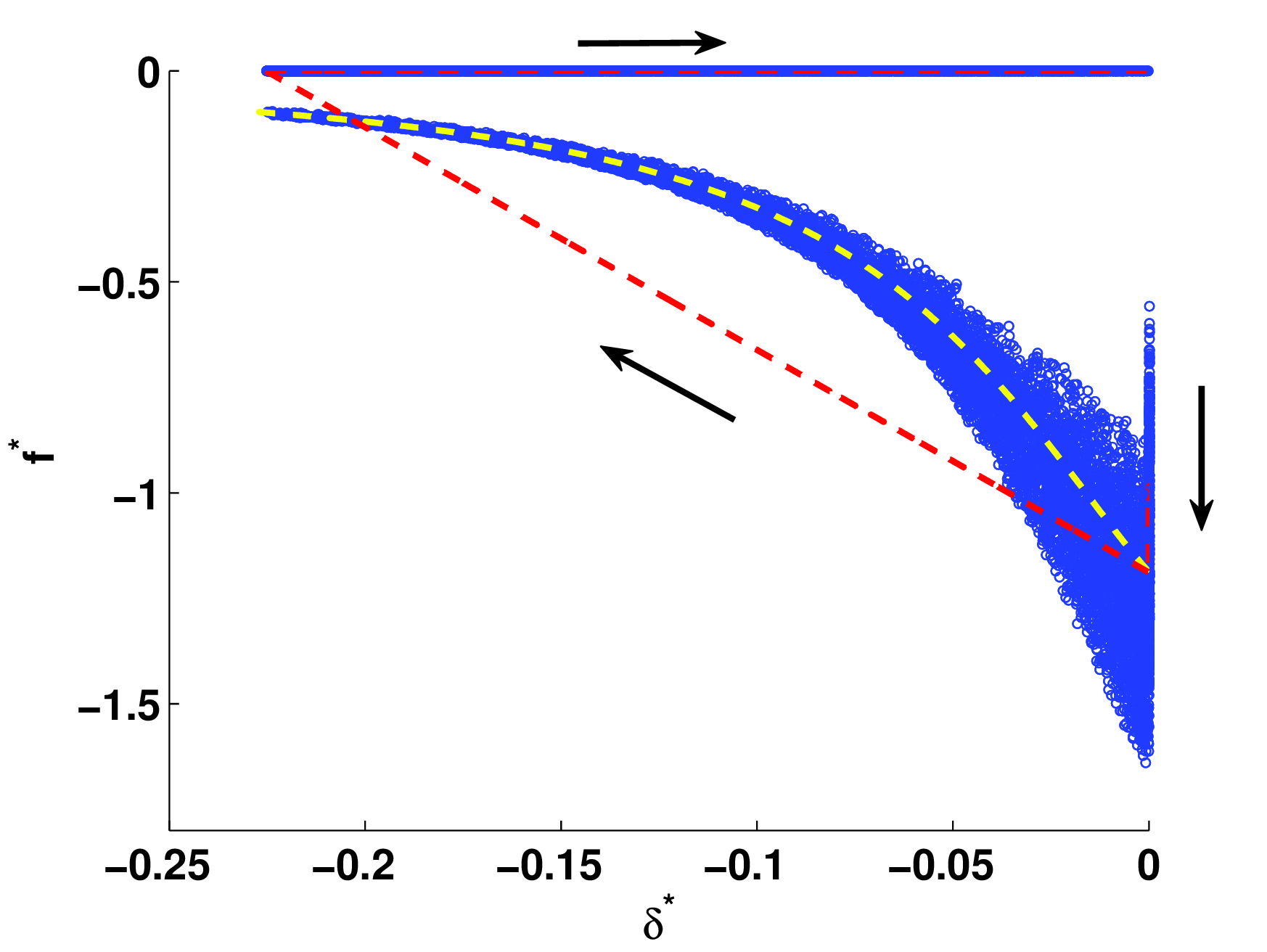}
   }%
  \end{center}
  \caption{Scaled adhesive force $f^*$ (${f_a}^*$ for linear adhesive model and ${f_c}^*$ for liquid bridge model) as a function of ${\delta}^*$ for the linear irreversible contact model (red), compared with the liquid bridge model (blue), for equal maximum force and equal interaction distance. The yellow line represents the force for the liquid bridge contact model for mean particle diameter $d_p$ as a function of ${\delta}^*$. The arrow shows the loading and the unloading directions for the short-range forces.}
  \label{fig:analogous_cohesive}
\end{figure}

The key parameters that define the cohesive force of a linear irreversible contact model are the maximum adhesive force and the adhesive stiffness, see Eq. \eqref{f_lin1}. Several simulations have been run for the linear irreversible contact model in the same numerical set-up with the same maximum adhesive force as used in the liquid bridge model ${(({f_a}^{max})}_{liq} = {({f_a}^{max})}_{lin})$ and adhesive stiffness that would result in the same interaction range for different liquid bridge volumes for different surface tension of liquid. The force-overlap for contacts with ${\delta}^* < 0$ for the two comparable contact models with equal interaction distance are shown in figure \ref{fig:analogous_cohesive}. The adhesive stiffnesses that are equivalent to the liquid bridge volumes as given by Eq. \eqref{vol_par} for surface tension $\gamma = 0.020$ Nm$^{-1}$ for equal interaction distance are given by:
\begin{multline} \label{adhstiff_par1}
k_{a} \ \in \ \left\{0.21\right., \ 0.26,  \  0.41, \ 0.46, \\
 0.56,  \ 0.69,  \   0.88,  \  1.11, \ \left. \infty\right\} \  \textnormal{Nm}^{-1}
\end{multline}
The results for the steady state cohesion $c^*$, as scaled by ${{\gamma}^*}$ for the liquid bridge model and the linear irreversible model are shown in figure \ref{fig:liq_lin_3}. The results are not really analogous as seen from the figure as the intercepts for the fitting lines of the two models are different, while they are parallel. The fitting parameters for the relation:
\begin{equation}
\frac{{c}^* - {c_0}^*}{{{\gamma}^*}} = g{S_c}^* + h
\label{eq:lin_rupture}
\end{equation}
are $g$ = 2.1716 and $h$ $\approx$ 0 for the liquid bridge contact model, $g$ = 2.0984 and $h$ = 0.2226 for the linear irreversible contact model.

So for a given liquid bridge volume and a given surface tension of liquid, the linear irreversible contact model with the same maximum force and same interaction distance has a higher cohesion. 

\subsection{Equal maximum force and adhesive energy}
\begin{figure}[!h]
  \begin{center}
  {%
	\includegraphics[width=\columnwidth]{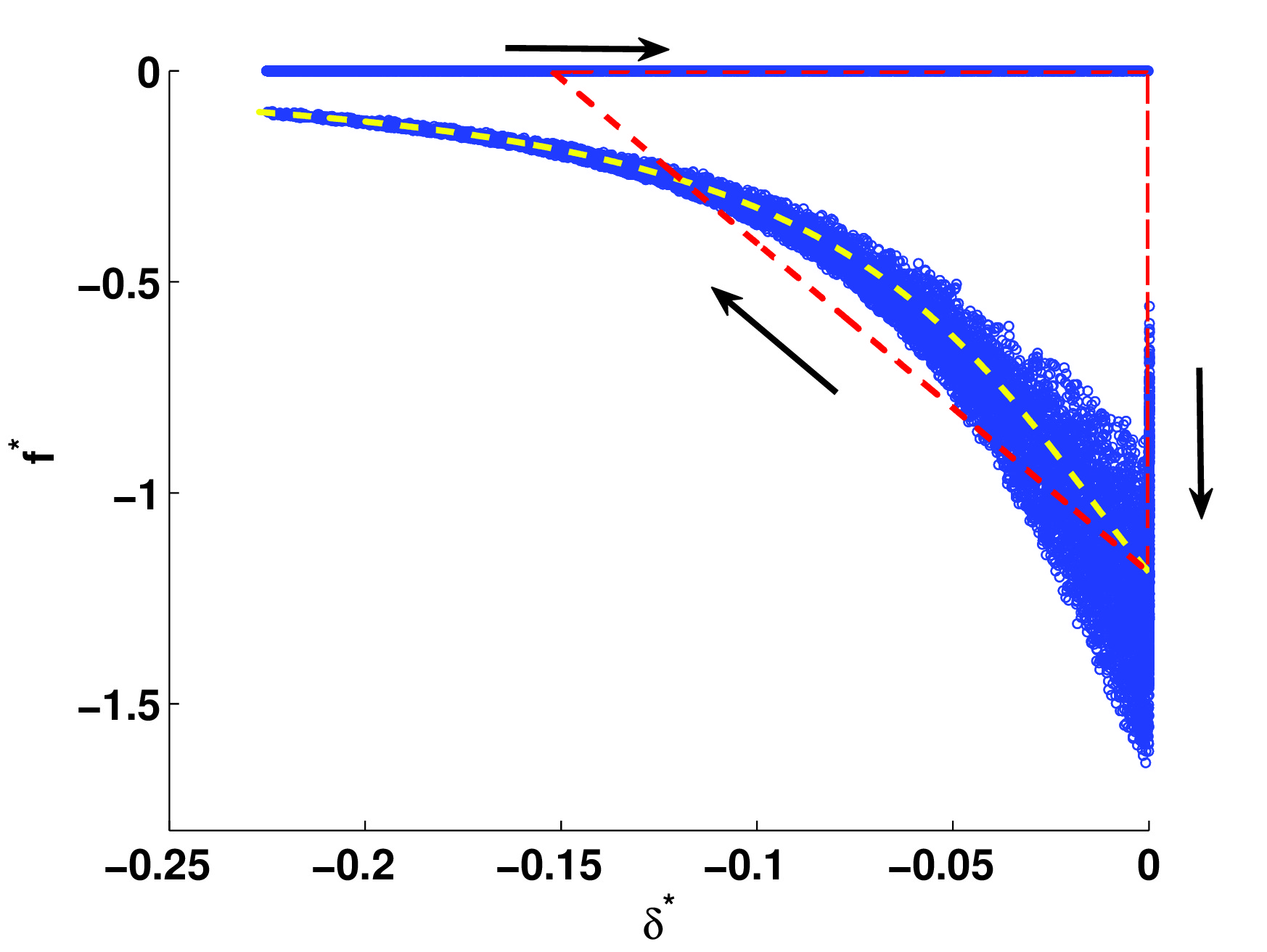}
   }%
  \end{center}
  \caption{Scaled adhesive force $f^*$ (${f_a}^*$ for linear adhesive model and ${f_c}^*$ for liquid bridge model) as a function of ${\delta}^*$ for linear the irreversible contact model (red), compared with the liquid bridge model (blue), for equal maximum force and equal adhesive energy dissipated per contact. The yellow line represents the force for the liquid bridge contact model for mean particle diameter $d_p$ as a function of ${\delta}^*$. The arrow shows the loading and the unloading directions for the short-range forces.}
  \label{fig:analogous1_cohesive}
\end{figure}

Equal maximum force and interaction distance was discussed in Sec. \ref{sec:analogous}, but here the steady state cohesion for the two models with an equal maximum adhesive force and equal adhesive energy $E^*$ are considered. The adhesive energy for a given contact model is obtained by the total area under the force-overlap distribution, see figure \ref{fig:analogous1_cohesive}. A linear model analogous to the liquid bridge contact model is obtained with the equal maximum force with surface tension $\gamma = 0.020$ Nm$^{-1}$ and the adhesive stiffness adjusted to have the equal adhesive energy:
\begin{multline}\label{adhstiff_par}
k_{a} \ \in \ \left\{0.25, \ 0.29\right., \ 0.39,  \  0.74, \ 0.84, \\
 1.10,  \ 1.49,  \   2.11,  \  2.95, \  \left.\infty\right\} \  \textnormal{Nm}^{-1}
\end{multline}
\begin{figure}[!h]
  \begin{center}
  {%
	\includegraphics[width=\columnwidth]{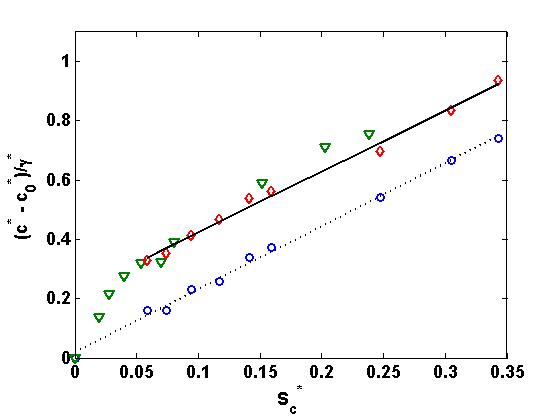}
   }%
  \end{center}
  \caption{$\frac{{c}^* - {c_0}^*}{{{\gamma}^*}}$ as function of ${S_c}^*$ for the liquid bridge model (blue) and the linear irreversible model with equal interaction distance (red) and equal adhesive energy dissipated per contact (green) for $\gamma = 0.020$ Nm$^{-1}$. The dotted and the solid lines represent the fitting function given by Eq. \eqref{eq:lin_rupture}.}
  \label{fig:liq_lin_3}
\end{figure}
\par
The force-overlap for contacts with ${\delta}^* < 0$ for the two comparable contact models with equal adhesive energy are shown in figure \ref{fig:analogous1_cohesive}. Figure \ref{fig:liq_lin_3} shows the dependence of $\frac{{c}^* - {c_0}^*}{{{{\gamma}^*}}}$ on rupture distance ${S_c}^*$ for the liquid bridge model (blue), compared with the two cases of the linear irreversible contact model with equal interaction distance (red) and equal adhesive energy dissipated per contact (green).  The linear irreversible model with equal energy has a lower interaction distance. The functional behavior of the steady state cohesion using the linear irreversible contact model for small interaction range can be understood from this. As observed from figure \ref{fig:liq_lin_3}, the cohesion is a non-linearly dependent on the rupture distance ${S_c}^*$ at low interaction distance and becomes linear for higher range. 
\par
\begin{figure}[!h]
  \begin{center}
  {%
	\includegraphics[width=\columnwidth]{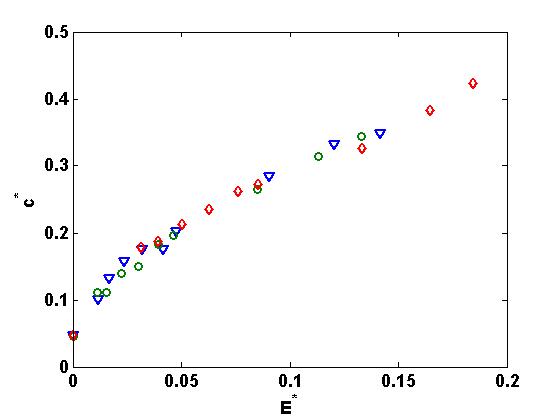}
   }%
  \end{center}
  \caption{${c}^*$ as a function of ${E}^*$ for the liquid bridge model (blue) and the linear irreversible model with equal interaction distance (red) and equal adhesive energy dissipated per contact (green) for $\gamma$ = 0.020 Nm$^{-1}$.}
  \label{fig:liq_lin_2}
\end{figure}

Figure \ref{fig:liq_lin_2} shows the dependence of steady state cohesion on total adhesive energy for the liquid bridge model, compared with the two cases of linear irreversible contact model with equal interaction distance (red) and equal adhesive energy dissipated per contact (green). As seen from the figure, for a given maximum force which is determined by the surface tension of the liquid, the steady state cohesion $c^*$ is equal for the the liquid bridge model and the linear irreversible model with equal energy. The steady state cohesion for the linear irreversible model with equal interaction distance is higher as it has higher adhesive energy than the liquid bridge model. However, all the data for the three cases as explained above collapse and functionally behave the same.

\subsection{Different maximum force for the two contact models}
\begin{figure}[!h]
  \begin{center}
  {%
	\includegraphics[width=\columnwidth]{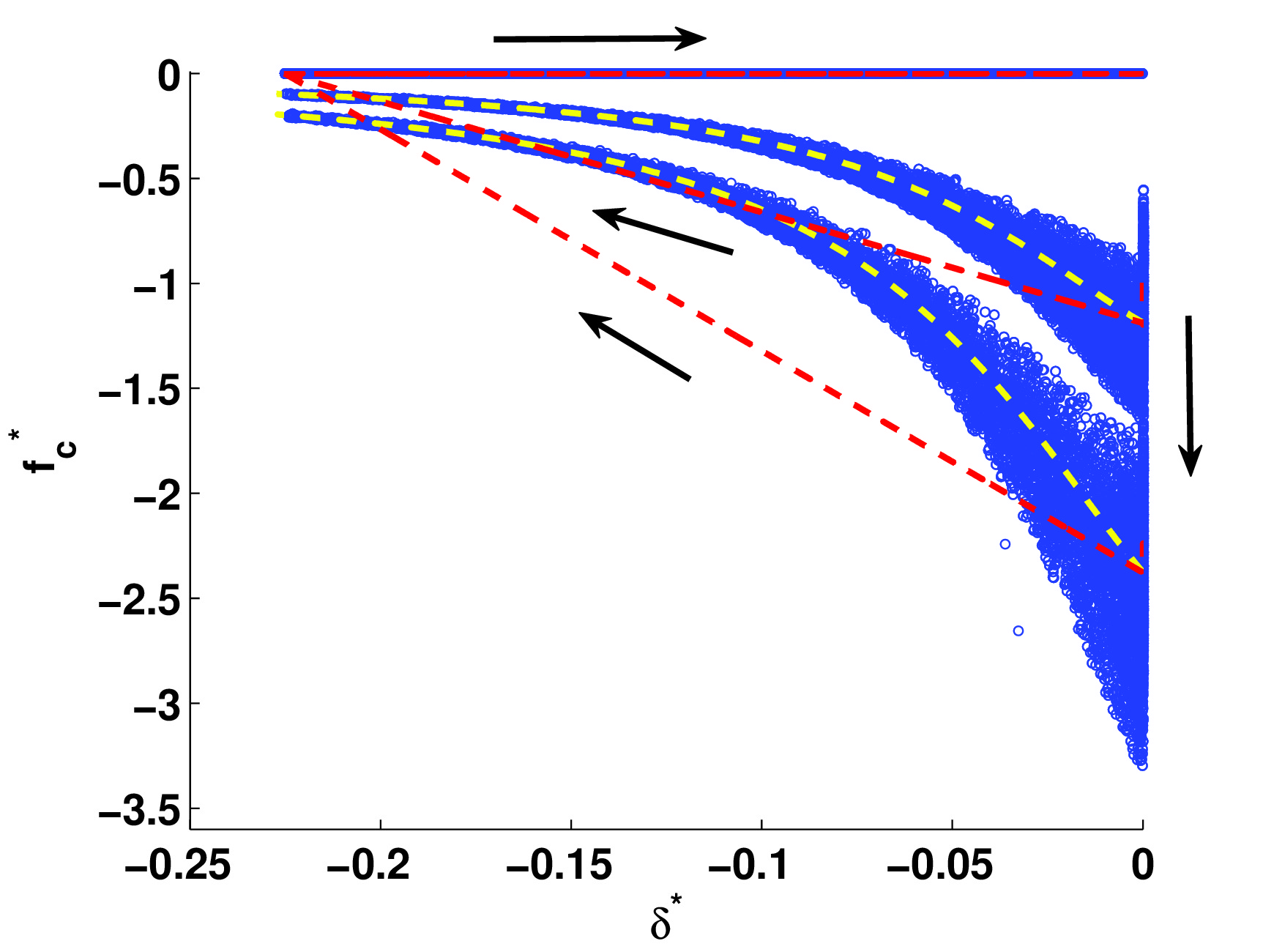}
   }%
  \end{center}
  \caption{Scaled adhesive force $f^*$ (${f_a}^*$ for linear adhesive model and ${f_c}^*$ for liquid bridge model) as a function of ${\delta}^*$ for the linear irreversible contact model (red), compared with the liquid bridge model (blue) for different maximum force and equal interaction distance. The yellow lines represent the force for the liquid bridge contact model for mean particle diameter $d_p$ as a function of ${\delta}^*$. The arrow shows the loading and the unloading directions for the short-range forces.}
  \label{fig:analogous3_cohesive}
\end{figure}

In the earlier subsections, results show that for a given maximum force the steady state cohesion for the two contact models functionally behave the same under equal force and equal energy conditions. To study the functional form for the two models under different maximum force conditions, we compare the macroscopic behavior of the linear model to the liquid bridge model results for different surface tension. Linear model simulations equivalent to surface tension 0.040 Nm$^{-1}$ and 0.060 Nm$^{-1}$ are run with an equivalent adhesive stiffness 2 times and 3 times of that given by Eq. \eqref{adhstiff_par1} keeping the interaction distance the same. Figure \ref{fig:analogous3_cohesive} shows a comparison of the force-overlap for the two contact models for surface tension of liquid 0.020 Nm$^{-1}$ and 0.040 Nm$^{-1}$. 
\begin{figure}[!h]
		
	\includegraphics[width=\columnwidth]{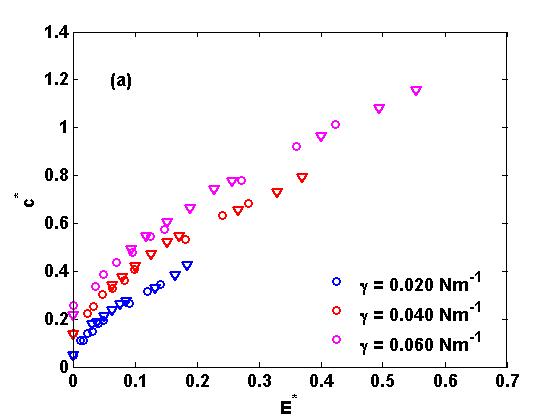}

	\includegraphics[width=\columnwidth]{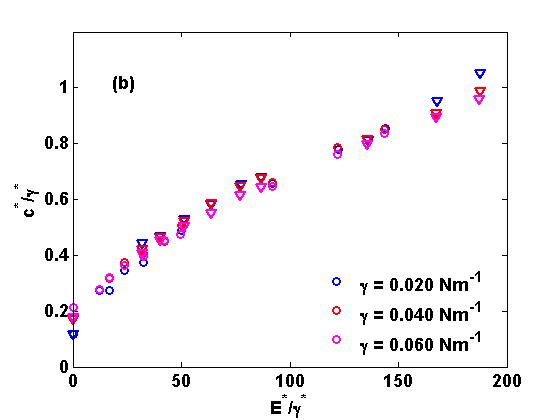}

\caption{(a) ${c}^*$ as a function of ${E}^*$ for different surface tension of liquid. (b)$\frac{{c}^*}{{{\gamma}^*}}$ as a function of  $\frac{{E}^*}{{{\gamma}^*}}$ for different surface tension of liquid as compared with the linear irreversible model. Different symbols denote $\circ$ liquid bridge model and $\nabla$ linear irreversible model. }
\label{fig:energy1}
\end{figure}
\par
Figure \ref{fig:energy1}(a) shows the dependence of steady state cohesion on the adhesive energy dissipated by the particles per contact for different ${f_a}^{max}$ for the liquid bridge model and the linear model. For the same energy dissipated per contact, a higher surface tension of the liquid results in a higher macroscopic cohesion. Figure \ref{fig:energy1}(b) shows that $\frac{{c}^*}{{{\gamma}^*}}$ is a function of $\frac{{E}^*}{{{\gamma}^*}}$ for a given surface tension, or maximum force.
\section{Conclusion}
We observed a correlation between the steady state cohesion and the microscopic parameters of the liquid bridge model. The micro-parameters are the liquid bridge volume, the liquid surface tension, the contact angle (which was kept constant) and the size of particles (i.e. curvature, which was also not varied). A detailed study of the effect of liquid bridge volume and surface tension of liquid is done in this paper. These microscopic parameters control the cohesion in wet granular materials in different ways. We found that the steady state cohesion of the system is proportional to the maximum adhesive force, which was varied by modifying the surface tension. On the other hand, we found that the steady state cohesion is also linearly dependent on the maximum interacting distance between the particles, which depends on the volume of the liquid bridge. From these results we have obtained a good micro-macro correlation between the steady state cohesion and the microscopic parameters studied. 
\par
We analyzed the effect of cohesion on the wall torque required to rotate the system at a given rate. The torque and the steady state cohesion of the system are proportional and show similar linear dependence on the microscopic parameters.
\par

Finally, an analogy was established between the liquid bridge model and a simpler linear irreversible contact model; even though these two models have different micro-macro correlations, the steady state cohesion for the two models are the same if the maximum force and the total adhesive energy dissipated per contact for the two models are matched, irrespective of the shape of the attractive force function acting between the particles. In this way one can always replace a non-linear liquid bridge force by a simpler, faster to compute linear one and obtain identical macroscopic properties in less computational time. 
\par

Furthermore, results for the two types of contact models with equal energy and different magnitude of maximum force show that they have different steady state cohesion. Therefore, adhesive energy is not the sole microscopic condition for the two contact models to have same steady state cohesion. Instead, we observe that both adhesive energy and cohesion scale linearly with the maximum adhesive force. In this way we can determine the steady state cohesion from the two microscopic parameters, the adhesive energy and the maximum force.
\par

In this paper our study was focused on the micro-macro correlations and comparing different contact models. It would be interesting to also study the forces and their probability distributions for wet cohesive systems. Future studies will aim at understanding the microscopic origin and dynamics of the contacts and liquid bridges throughout the force network(s) and also the directional statistics of the inter-particle forces inside a shear band. The effect of liquid migration on the macro properties and a continuum description for a similar model will be studied in the near future.

\begin{acknowledgements}
We acknowledge our financial support through STW project 12272 "Hydrodynamic theory of wet particle systems: Modeling, simulation and validation based on microscopic and macroscopic description."

\end{acknowledgements}

\bibliographystyle{spphys}       


\end{document}